\documentclass[12pt]{article}
\usepackage{aasms4}
\usepackage{tighten}
\usepackage{natbib}
\usepackage{psfig}
\begin{document}


\newcommand{\bq}{\begin{equation}}
\newcommand{\eq}{\end{equation}}  

\bibliographystyle{natbib-apj}

\title{Snapshot Distances to Type Ia Supernovae -- All in ``One''
Night's Work} 

\author{Adam G. Riess\altaffilmark{1}, Peter Nugent\altaffilmark{2}, \\
Alexei V. Filippenko\altaffilmark{1,3}, Robert P. Kirshner\altaffilmark{4},
and Saul Perlmutter\altaffilmark{2,3}}
 
\altaffiltext{1}{Department of Astronomy, University of California,
Berkeley, CA 94720-3411; ariess, alex@astro.berkeley.edu}

\altaffiltext{2}{Lawrence Berkeley National Laboratory, Berkeley, CA
94720; penugent, saul@lbl.gov}

\altaffiltext{3}{Center for Particle Astrophysics, University of
California, Berkeley, CA, 94720} 

\altaffiltext{4}{Harvard-Smithsonian Center for Astrophysics,
Cambridge, MA 02138; kirshner@cfa.harvard.edu}
 
\begin{abstract}

We present an empirical method which measures the distance to a Type
Ia supernova (SN~Ia) with a precision of $\sim$ 10\% from a single
night's data.  This method measures the supernova's age and
luminosity/light-curve parameter from a spectrum, and the extinction
and distance from an apparent magnitude and color.  We are able to
verify the precision of this method from error propagation
calculations, Monte Carlo simulations of well-sampled SNe~Ia, and the
Hubble diagram of scarcely observed supernovae.  With the reduction in
telescope time needed, this method is three to four times more
efficient for measuring cosmological parameters than conventional
light-curve based distance estimates.

\end{abstract}

 
\section{Introduction}

The explosion of a Type Ia supernova (SN~Ia) is a catastrophic
phenomenon veiled in layers of complexity.  Recent efforts to monitor
these events have led to an increased ability to predict, if not fully
understand, the stages of SN Ia evolution.

The model for the photometric history of SNe~Ia has been refined from
a homogeneous description \cite[]{brunophd,branch_miller,ts95} to one
which characterizes a relation between peak luminosity and light-curve
shape \cite[]{philm15,hametal95,hametal96,rpk95,kim_stretch98}. The
slower, broader light-curves are intrinsically brighter at peak than
the faster, narrower light-curves. Recognizing and exploiting such
relations has led to a renaissance in the use of SNe~Ia as
extragalactic distance indicators.  Extending luminosity/light-curve
relations to multiple passbands separates the competing effects of
dust, intrinsic differences, and distance on the light of SNe~Ia
\cite[]{rpk96}.  Distances with 5-10\% uncertainty can be obtained
using the light-curve shapes of well-observed supernovae.

The optical spectra of SNe~Ia are rich in information [see
\cite{fil97} for a review].  Many of the elements synthesized and
ejected in the explosion have been identified despite the blending of
their high-velocity profiles \cite[]{bran81b,nugphd}.  In addition,
the relative strengths of some spectral features have been shown to
correlate with SN~Ia peak luminosity \cite[]{nugseq95}.  As the
supernova evolves, predictable casts of features appear and disappear,
illuminated by the photosphere's recession through the synthesized
layers.  The temporal evolution of these features is sufficiently
reliable to be used as a clock to determine the current age of a SN~Ia
to a precision of 1-2 days \cite[]{mink39,riess_age97}.

Unfortunately, supernovae occur without warning, making it difficult
to collect the observations necessary to measure their distances.
Observing an unscheduled event in up to four filters many times over
the course of $\sim$ 100 days is a time consuming and logistically
formidable task. The observing record for a typical SN~Ia is quite
fragmentary.  Following this process, it will be many years of work to
gather the number of SN~Ia distances necessary to put strong limits on
cosmological parameters. Even at high redshifts ($z \geq 0.3$), where
a strategy for batch detections of multiple supernovae has made it
possible to schedule supernova discoveries and their follow up
\cite[]{perl97}, difficulties arise. Since these observations require
the largest telescopes, the light-curves are typically more poorly
sampled than the nearby ones caught at a similar phase.

Yet, from a single night's observations, a SN~Ia's spectrum and
photometric magnitudes can reveal its age, intrinsic luminosity,
extinction, and apparent brightness.  From this information one can
estimate the distance to a supernova without further observations
(except for the possible need of a galaxy image to subtract the host's
light).

Here we explore this possibility with two independent sets of SNe~Ia.
We describe this technique in \S 2 and its expected uncertainty in \S
3.  In \S 4 we apply it to randomly selected snapshots of extensively
observed SNe~Ia to determine the precision of such distance estimates.
In \S 5 we construct the Hubble diagram of ``cast-off'' SNe~Ia:
objects which were observed only once or a few times.  We extend the
application of this method in \S 6 to SNe~Ia with $0.2 \leq z \leq
0.83$.  In \S 7 we discuss variations of this technique and its
leverage on estimating cosmological parameters.

\section{A Snapshot Distance}

\begin{figure}[p]
\psfig{file=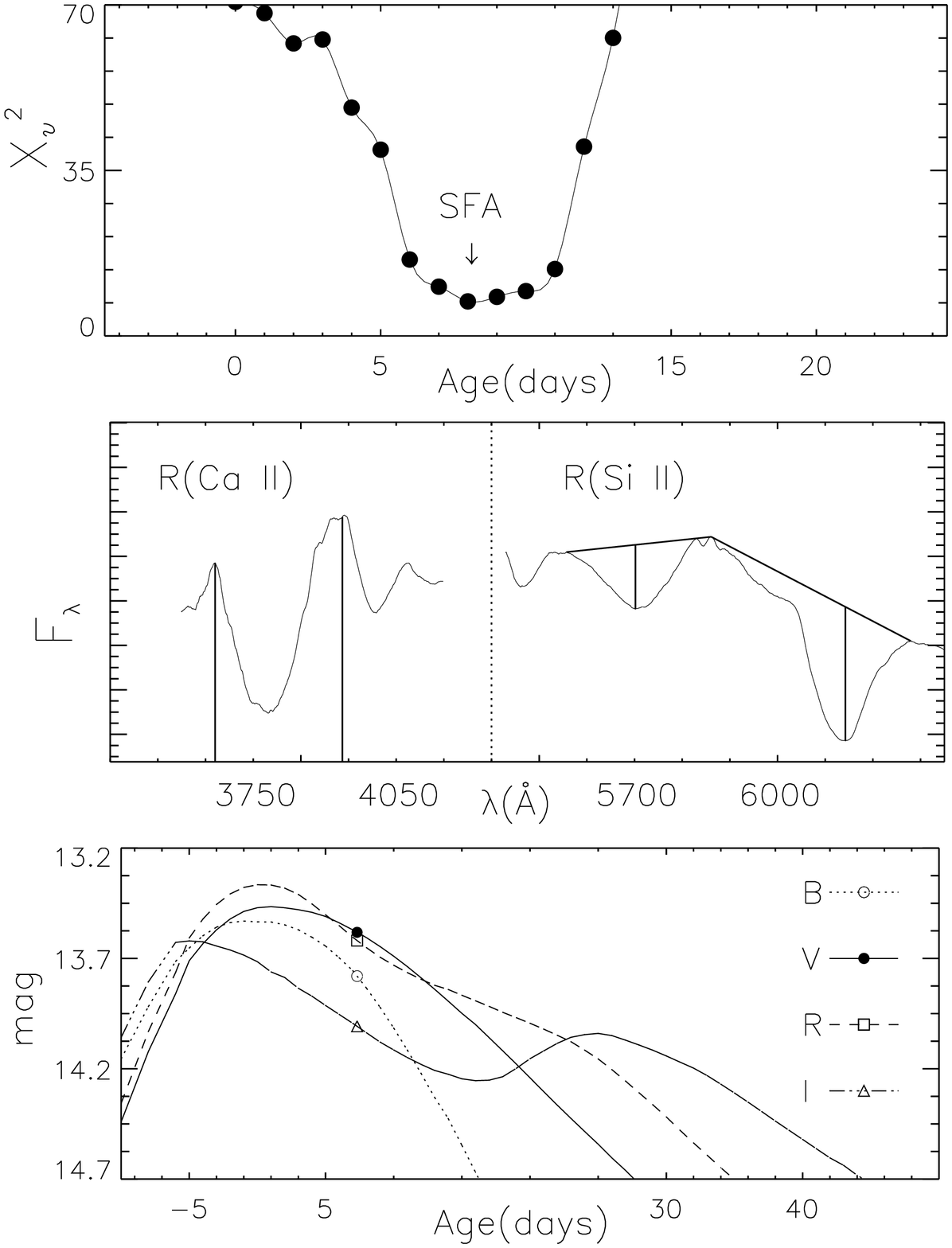,height=6.0in,width=6.0in,angle=0}
\caption{An example of the snapshot distance method for SNe~Ia as
applied to SN 1995D.  (a) By measuring a SN~Ia spectrum's
goodness-of-fit to a database of SNe~Ia spectra as a function of
light-curve age we can estimate the most likely age of the spectrum.
(b) The ratio of \ion{Si}{2} absorption troughs at 5180 and 6150~\AA\
and the ratio of peaks associated with \ion{Ca}{2} at 3940 and
3970~\AA\ indicate the individual luminosity/light-curve parameter of
the SN~Ia (Appendix A). (c) Using the estimates of the age and
luminosity/light-curve parameter, a single epoch of multicolor
photometry constrains the extinction and the distance.\label{one}}
\end{figure}

All of the information necessary to estimate the distance to a SN~Ia
may be garnered from a single spectrum and epoch of photometric
magnitudes.  Below we describe how to determine a SN~Ia's age,
intrinsic luminosity, extinction, and distance from this glimpse. An
example of how this is done is illustrated in Figure~\ref{one}.

The absence or presence of features in a SN~Ia spectrum is a sensitive
indicator of its current age.  \cite{riess_age97} developed an
algorithm to measure the spectral feature age (SFA) of a SN~Ia by
comparing the goodness-of-fit of its spectrum to a database of SNe~Ia
spectra of known age.  For high signal-to-noise ratio spectra ($S/N
\ge 40$), the age uncertainty is $\sim$ 1.4 days.  Given a single SN
Ia spectrum, we estimate the SFA in this way (Figure~\ref{one}a).
With two or more spectra, we can improve the age estimate by adding
the individual goodness-of-fit curves in the time frame of the
supernova.

Having determined the age of the spectral epoch(s), we can estimate
the intrinsic luminosity of the SN~Ia by measuring the ratio of
specific spectral features.  \cite{nugseq95} have shown that at
maximum light the ratios of the depths of the \ion{Si}{2} absorption
features at 5180 and 6150~\AA\ [$\cal{R}$(\ion{Si}{2})] and the
heights of features at 3940 and 3970~\AA\ associated with \ion{Ca}{2}
[$\cal{R}$(\ion{Ca}{2})], are strongly correlated with the SN~Ia peak
luminosity and light-curve shape.  Here we extend that method by
correlating a luminosity/light-curve parameter with these ratios
measured over the range in supernova age of six days before maximum to
twenty days after maximum.  For the spectral database considered by
\cite{riess_age97} and spectra of SNe~Ia considered by \cite{nugseq95}
and \cite{phillips97} -- a total of $\sim 200$ spectra -- we measured
$\cal{R}$(\ion{Si}{2}) and $\cal{R}$(\ion{Ca}{2}) as prescribed by
\cite{nugseq95}.  For this same set of SNe~Ia, we measured the
luminosity/light-curve parameter from their multicolor light-curve
shapes (MLCS) \cite[]{rpk96}.  We then derived the linear
relationships between the MLCS luminosity/light-curve parameters and
the spectral ratios as a function of light-curve age.  These relations
are described in detail in Appendix A.
 
For a given SN~Ia spectrum we measure $\cal{R}$(\ion{Si}{2}) and
$\cal{R}$(\ion{Ca}{2}), as shown in Figure~\ref{one}b.  We then
estimate its luminosity/light-curve parameter from the spectral ratio
relation valid for the spectrum's SFA. If there is more than one
spectral ratio or more than one spectrum we average the estimates of
the luminosity/light-curve parameter weighted by the dispersion of the
relations given in Appendix A.

The luminosity/light-curve shape parameter identifies the expected
photometric history for the SN~Ia in up to four passbands: $B,V,R,$
and $I$ \cite[]{rpk96}.  The age estimate determines when in the
course of SN~Ia history a photometric epoch was observed.  For SNe~Ia
with $z \geq 0.01$ we correct for the effect of (1+$z$) time dilation
on the {\it difference} in time between the spectral and photometric
observations.  We also include a K-correction on the supernova light
as described by \cite{hametal93} and for those at higher redshifts ($z
\geq 0.2$) as described by \cite{kim_kcorr96}. By placing the observed
magnitudes at the appropriate time on the light-curves (with their
shapes determined from the spectral ratios), we determine the apparent
distance moduli (Figure~\ref{one}c).  Because of the reddening and
absorbing properties of interstellar dust, the distance moduli of the
shorter wavelength bands are expected to be equal to or greater than
those of longer wavelength bands.  These differences are the color
excesses caused by intervening dust.  If a red color excess is
measured, we can use the conventional interstellar extinction law
\cite[]{savage,rpk96dust} to correct the supernova light for
interstellar extinction.

Though it may seem surprising, the natural consequence of the results
of Riess et al. (1997a), Riess, Press, \& Kirshner (1996b), and Nugent
et al. (1995) is that the distance to a SN Ia can be measured from a
single night's observations.  These works have established the tools
which, when combined, reveal the predictive powers of a single epoch
of SN Ia data.  In the following sections we seek to answer the
question of how well such limited data can constrain the distance to a
SN Ia.

\section{Theory of Errors}

We can estimate the {\it expected} precision of these snapshot
distances by propagating all known sources of error which affect the
distance estimate.

Apparent distance moduli may be calculated from $V$ and $B$ apparent
magnitudes, $m_V$ and $m_B$, measured when a SN~Ia is $\tau$ days
after $B$ maximum ($\tau$ may be negative), as

\bq 
\mu_v=m_v-M_v(\tau)-R_v(\tau)\Delta, 
\label{mu_v}
\eq
\bq
\mu_b=m_b-M_b(\tau)-R_b(\tau)\Delta, 
\label{mu_b}
\eq

\noindent where $M_v$ and $M_b$ are the broad-band absolute magnitudes
of a ``fiducial'' SN~Ia, and $\Delta\equiv M_v-M_v(fiducial)$ is the
difference in maximum ($\tau=0$) $V$ luminosity between the fiducial
SN~Ia and any other; $\Delta$ is the luminosity/light-curve parameter
for the MLCS method discussed in \S 2.  $R_v(\tau)$ and $R_b(\tau)$
are proportionality constants which, when multiplied by $\Delta$,
describe the luminosities of individual SNe~Ia as a function of their
age \cite[]{rpk96}.

The quantity of interest is the {\it extinction-free distance},
$\mu_0$, 

\bq 
\mu_0 = \mu_v-3.1(\mu_b-\mu_v), 
\label{mu_0}
\eq 

\noindent where we assume the Galactic extinction law is valid for SNe
Ia (Savage \& Mathis 1979; Riess, Press, \& Kirshner 1996b).  The
expected error in the extinction-free distance, $\sigma_{\mu_0}$, is
given by substituting equations~(\ref{mu_v}) and (\ref{mu_b}) into
(\ref{mu_0}) and applying error propagation:

\begin{eqnarray}
\label{sig1}
\sigma_{\mu_0}^2 & = & 4.1^2\sigma_{m_v}^2 + 3.1^2\sigma_{m_b}^2\\\nonumber
         & + & \sigma_{\tau}^2 \{ 3.1 \frac {\partial M_b(\tau)} 
           {\partial \tau} - 4.1 \frac {\partial M_v(\tau)} 
           {\partial \tau}\\\nonumber
         & + &  \frac{\partial \Delta} {\partial \tau}
           [3.1 R_b(\tau) - 4.1 R_v(\tau)] + \Delta [3.1 \frac
           {\partial R_b(\tau)} {\partial \tau} - 4.1 \frac {\partial
           R_v(\tau)} {\partial \tau}]\}^2\\\nonumber
         & + & \sigma_{\Delta}^2 [3.1 R_b(\tau) - 4.1 R_v(\tau)]^2. 
\end{eqnarray}

Both terms on the third line of equation~(\ref{sig1}) contribute an
error of less than 0.02 mag to the distance estimate because $R_b(\tau)$,
$R_v(\tau)$, and the value of $\Delta$ inferred from the spectral
ratios are slowly varying functions of supernova age, $\tau$. The
remaining error is
\begin{eqnarray}
\label{sig2}
\sigma_{\mu_0}^2 & \approx & 4.1^2\sigma_{m_v}^2 + 
	3.1^2\sigma_{m_b}^2\\\nonumber 
	& + & \sigma_{\tau}^2 (\frac{-\partial M_v(\tau)}{\partial \tau} +
         3.1\frac{\partial M_{b-v}(\tau)}{\partial \tau})^2\\\nonumber 
	& + & \sigma_{\Delta}^2 (-R_v(\tau) +
         3.1R_{b-v}(\tau))^2,
\end{eqnarray}

\noindent where we define $M_{b-v} \equiv M_b-M_v$ and $R_{b-v} \equiv
R_b-R_v$ to separate SN Ia luminosity from color.

The first line in equation~(\ref{sig2}) depends only on the errors in
the photometric magnitudes.  For a likely photometric error of 0.02
mag for nearby ($z < 0.1$) SNe Ia, this error contributes 0.10 mag to
the distance uncertainty.  The remaining terms in
equation~(\ref{sig2}) are fortuitously small, as follows.

\begin{figure}[p]
\psfig{file=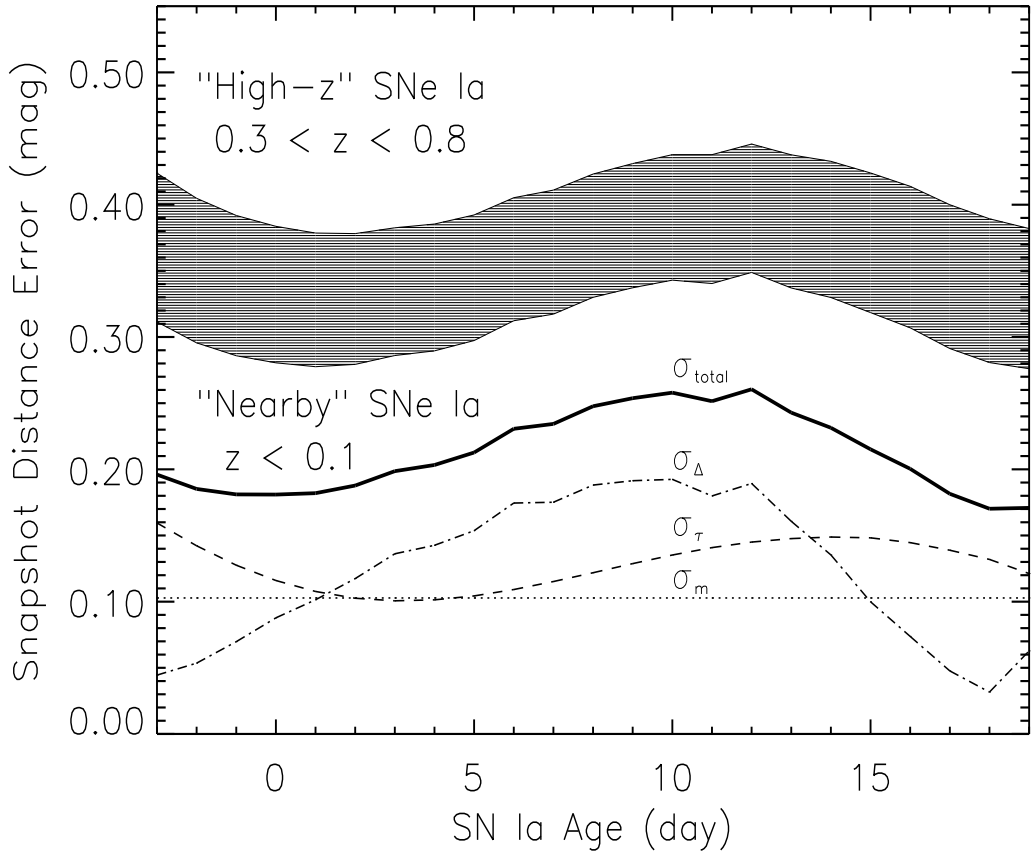,height=5.0in,width=6.0in,angle=90}
\caption{Expected snapshot distance uncertainty as a function of age
for nearby and distant SNe~Ia.  With SFA $1\sigma$ errors of 1.4 days
and photometry $1\sigma$ errors of 2\% for SNe~Ia with $z$ $\leq$ 0.1,
equation~(\ref{sig2}) predicts a $1\sigma$ distance uncertainty of
$\sim$ 10\% from a single spectral and photometric epoch.  The
distance error resulting from the photometry uncertainty ($\sigma_m$),
the SFA uncertainty ($\sigma_{\tau}$), and the luminosity parameter
uncertainty ($\sigma_\Delta$) are indicated.  For SNe~Ia with $ 0.3
\leq z \leq 0.8$ the increase in SFA $1\sigma$ errors to 2.0-2.5 days
and $1\sigma$ photometry errors to 4\% to 6\% decreases the $1\sigma$
snapshot distance precision to 15\% to 20\% .\label{err}}
\end{figure}

The second line in equation~(\ref{sig2}) depends on the uncertainty,
$\sigma_{\tau}$, in measuring the age of a SN~Ia at the time of the
photometric epoch. An incorrect estimate of the age has two effects on
the distance estimate which are compensatory.  For example, on the
post-maximum side of the light-curve, an underestimate of the age
would result in an overestimate of the supernova's current luminosity
and the prediction of a distance which is {\it too far} by the amount
$\sigma_{\tau}( \partial M_v(\tau) / \partial \tau)$.  But this
underestimate of the age would also result in a prediction that the
SN~Ia is currently bluer and therefore more extinguished for its
observed color.  This overestimate of the extinction would give a
prediction for the distance which is {\it too close} by the amount
$3.1\sigma_{\tau}( \partial M_{b-v}(\tau) / \partial \tau)$.  For
expected age uncertainties of 1.4 days \cite[]{riess_age97} over the
first three weeks of typical SN~Ia light-curves \cite[]{rpk96}, the
sum of these compensating errors contributes a distance uncertainty of
0.10 to 0.15 mag (see Figure~\ref{err}).  However, on the pre-maximum
side of the light-curve this compensation does not occur.  If spectra
are obtained which indicate the SN~Ia is pre-maximum, it is desirable
to obtain additional photometry no earlier than 3 days before maximum
to take full advantage of the above effect which reduces the distance
error.

The last line in equation~(\ref{sig2}) which rises with the
uncertainty, $\sigma_{\Delta}$, in the estimate of the
luminosity/light-curve parameter $\Delta$, is also limited by
compensating effects.  An incorrect estimate of the
luminosity/light-curve parameter leads to an erroneous estimate of the
luminosity and color and hence of the extinction.  These effects have
a partially balancing influence on the distance estimate.  For
example, an overestimate of the luminosity by the amount $\delta$
would lead to a prediction that the SN~Ia is {\it farther} away by
$\delta R_v(\tau)$, but the subsequent prediction that the SN~Ia is
intrinsically bluer leads to an overestimate of the extinction, and
hence a prediction that the distance is {\it closer} by the amount
$\delta 3.1R_{b-v}(\tau)$.  For likely uncertainties in the estimate
of the luminosity/light-curve parameter, $\Delta$, of 0.20 to 0.25 mag
(see Appendix A), we find this term contributes a 0.03 to 0.18 mag
uncertainty to the distance estimate depending on the age of the
supernova (see Figure~\ref{err}).  This effect was discussed by
\cite{vdb95} and \cite{branch_vdb96} and may be summarized by noting
that a red supernova might be either intrinsically dim or extinguished
by dust.  The sizes of these effects on the distance are slightly
different \cite[]{rpk96}, but their {\it direction} is the same.

Combining all of these expected sources of error leads to the
predicted distance uncertainties in Figure~\ref{err}.  The
uncertainty, for a single spectrum and epoch of $B$ and $V$
photometry, ranges between 0.17 and 0.26 mag depending on the age of
the SN Ia.  For noisier data expected to accompany SNe~Ia at $z \geq
0.2$, this uncertainty will be greater (see \S 6).  Within the first
week after maximum, the distance precision is greatest because of the
slow rate of change of the light-curves as well as the availability of
both the \ion{Si}{2} ratio and the \ion{Ca}{2} ratio.  Procuring data
in three or four passbands and at multiple epochs will lead to modest
improvements to the distance precision.  Yet additional data are
invaluable as a guard against systematic errors in measuring the SNe
Ia apparent magnitudes and colors.  These errors can result from
cosmic rays hits, defective pixels, or poor background subtraction.

\section{Simulated Snapshots}

Although our ambition is to measure SNe~Ia distances from as little as
one spectrum and photometry epoch, we can test the feasibility of this
endeavor on more extensively monitored objects.  Our procedure is to
select, at random, one epoch of photometry and one spectrum from the
set of observations of SNe 1993ac, 1993ae, 1994M, 1994Q, 1994S, 1994T,
1995D, 1996C, and 1996X.  These SNe~Ia are in the range 2000
km~s$^{-1}$ $\leq$ $cz$ $\leq$ 15000 km~s$^{-1}$ and were extensively
observed by \cite{riessphd} and \cite{riess_data97}.

We calculated the distance for random ``snapshots'' of these SNe~Ia by
the procedure described in \S 2, and measured the dispersion of these
distances around the best-fit Hubble line.  We repeated this exercise
to build a distribution of dispersions.  With this ``Monte Carlo''
exercise, we can determine the statistical and systematic components
of error for snapshot distances.  Because of our inability to reliably
make use of $\cal{R}$(\ion{Si}{2}) and $\cal{R}$(\ion{Ca}{2}) earlier
than six days before maximum or later than twenty days after maximum
(see Appendix A), we restricted our selection to epochs within this
range.

The solid line in Figure~\ref{show} shows the distribution of Hubble
dispersions for simulated snapshot distances.  The most frequently
occurring dispersion was 0.21 mag in the distance modulus with a
standard deviation of $0.03$ mag, implying a remarkable precision of
$\sim$ 10\% in distance. For comparison, the original treatment of the
data using full light-curves produces a dispersion of 0.12 mag or
$\sim$ 6\% in distance (Riess, Press, \& Kirshner 1996a). The
accompanying distribution of Hubble constants peaks at $64.1 \pm 0.9$
km~s$^{-1}$ Mpc$^{-1}$, consistent with the value of $64.5$
km~s$^{-1}$ Mpc$^{-1}$ determined from MLCS distances for these same
nine SNe~Ia, implying that {\it no additional systematic bias} is
apparent in the snapshot distances relative to light-curve distances.
These uncertainties in the Hubble constant reflect only the small
point-to-point variance of the observed Hubble flow and do not account
for the much larger uncertainty in the determination of the SN Ia
absolute magnitude and the connection to the Cepheid distance scale.
These latter uncertainties limit the determination of the Hubble
constant to $\sim$ 10\% precision \cite[]{feast87,koch97,madore98}.
The snapshot distances do not provide an independent measurement of
the Hubble constant or of the distances to these SNe Ia.  Rather, our
distance indicator is estimating the distances which {\it would have
been} predicted by the MLCS method (RPK 1996a) if well-sampled
light-curves were available.

\begin{figure}[p]
\psfig{file=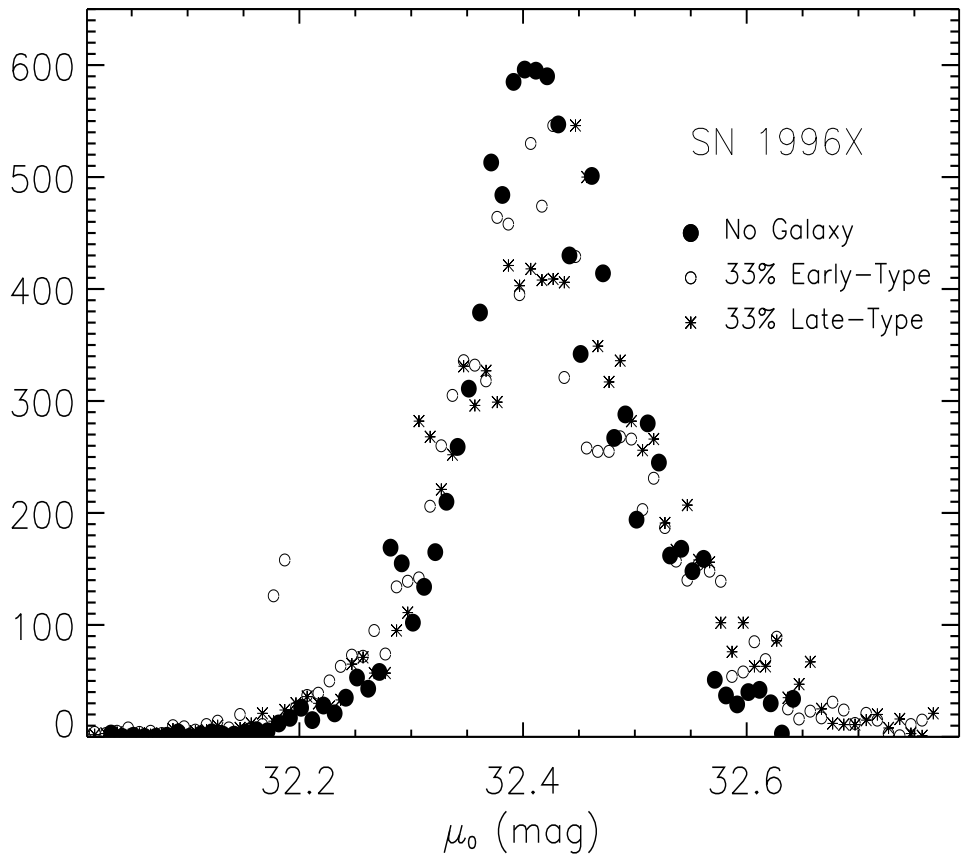,height=5.0in,width=6.0in,angle=90}
\caption{Distribution of snapshot method distances to a single SN Ia
(SN 1996X) with and without host galaxy contamination.  A single
spectrum and photometric epoch were randomly selected (not necessarily
the same date) from the available observations of SN 1996X to measure
the snapshot distances from different epochs.  As a separate exercise,
the spectra of SN 1996X were contaminated by early-type and late-type
galaxy light at the level of 33\% of the supernova brightness ($B$
band).\label{96X}}
\end{figure}

An alternate way to explore the variance of the snapshot distances is
to determine the distribution of snapshot distances for a single
well-observed object.  This is the same Monte Carlo exercise employed
above except we do not compare the simulated snapshot distances to the
distance predicted from the supernova's redshift.  Consequently, the
distribution of distances for an individual SN Ia is not broadened by
uncertainties in absolute photometry or by the peculiarities of a SN
Ia which might cause a systematic mis-estimate of its SFA or
luminosity parameter from {\it any} spectral epoch.  Rather, this
distribution shows the range of snapshot distances which would have
been estimated from a single spectrum and photometric observation
collected at {\it different epochs}.  For this exercise we used SN
1996X which had the greatest number of available spectral epochs (10)
and photometric epochs in four colors (12 per color), yielding over
200,000 combinations of possible snapshot data sets.  For each
randomly selected snapshot data set we estimated the snapshot distance
to SN 1996X; the resulting distribution has a dispersion of 0.10 mag
about the mean and is given by the solid symbols in Figure~\ref{96X}.

The light of the SN Ia host galaxy can contaminate the measured light
of the SN Ia if not properly subtracted (e.g., Boisseau \& Wheeler
1991).  Such galaxy contamination is commonly removed from supernova
photometry (e.g., Schmidt et al. 1998), but is not typically removed
from SNe Ia spectra.  We performed a test of the effects of host
galaxy contamination on the snapshot distances by repeating the above
exercise with SN 1996X using spectra contaminated by galaxy light.
Specifically, we added early-type and late-type galaxy light to the
spectra of SN 1996X in the amount of 33\% of the SN brightness ($B$
band).  The resulting distribution of snapshot distances derived from
these contaminated spectra is shown in Figure~\ref{96X}.  In general
we find a small decrease in the inferred SFA of 0.5 to 1.0 days and a
mild decrease of $\sim$ 15\% in the value of $\cal{R}$(\ion{Si}{2})
and $\cal{R}$(\ion{Ca}{2}).  Again, due to the compensating effects
described in \S 3, the affect on the inferred distances is
substantially reduced.  Late-type galaxy contamination on the SN Ia
spectra at this level increased the mean distance by 0.02 mag and
increased the dispersion from 0.10 to 0.11 mag.  Early-type
contamination decreased the mean distance by 0.01 mag and increased
the dispersion from 0.10 to 0.11 mag.

In \S 7 we discuss the random and systematic errors of distance
estimates from a single epoch of SN Ia observations which neglect the
effects of extinction and intrinsic luminosity differences.

Having estimated and simulated the precision of snapshot distances, we
now apply this technique to SNe~Ia which were sparingly observed.

\section{The Hubble Diagram of Snapshot Distances}

The primary result of \S 3 and \S 4 is that snapshot distance
estimates have uncertainties of $\sim$ 0.2 mag or 10\%. A convincing
demonstration of the distance precision available from single-epoch
observations can be made with an independent set of SNe~Ia which were
observed only one or a few times.  For such a set, no other means of
measuring the distance to the SNe~Ia is available.

We have searched for recent SNe~Ia which were observed only one to a
few times spectroscopically and photometrically.  Our selection
criteria were based on the feasibility of measuring a snapshot
distance and were as follows.

\begin{enumerate}

\item A spectroscopically confirmed SN~Ia which yields a reliable SFA
within twenty days after maximum light.

\item At least one epoch of CCD photometry available in at least two
passbands.

\item Recession velocity $cz \ge 1000$ km~s$^{-1}$.

\end{enumerate}

Condition 1 is necessary to be able to make a useful estimate of the
luminosity/light-curve parameter from the ratio of spectral features
as well as to be able to determine when along the SN~Ia light-curve a
photometric epoch was observed.  The measurement of a reliable SFA is
a {\it prerequisite} for determining the luminosity/light-curve
parameter from $\cal{R}$(\ion{Si}{2}) and $\cal{R}$(\ion{Ca}{2}) as
the use of these ratios requires knowledge of the age of a SN Ia
spectrum (Appendix A).  A current limitation of this method is posed
by two rare and unusual varieties of SNe~Ia, characterized by SN 1991T
and SN 1991bg \cite[]{fil91bg92,fil91bg,phil91t,bfn93}.  The presence
of atypical features as well as the absence of typical features in
such spectra hinders our ability to locate such objects within our
database of SNe~Ia temporal evolution.  Specifically, the SFA method
(Riess et al. 1997a) does not provide a robust indication of the age
of these atypical SNe Ia near maximum light.  Until more such objects
are observed, it will not be possible to measure their age reliably
from their spectral evolution.  With the addition of recent examples
SN 1995ac, SN 1997br, SN 1997cw, and SN 1997cn \cite[]{riess_data97}
(see also IAUC 6642, 6699 and 6661) to the literature, we expect even
this limitation to be lifted.  More than a week after maximum, spectra
of SN 1991T began to resemble typical SNe Ia.  At such times, SN 1991T
does provide reliable SFA's; the spectra 10 and 15 days after maximum
yield an SFA of 7.5 and 13.3 days after maximum, respectively
\cite[]{fil91bg92,phil91t}.  Consequently, SN 1991T would pass
condition 1 if spectroscopically observed more than 1 week after
maximum.

Condition 2 requires that photometry from at least two passbands be
available to measure the extinction from the color excess.  Although
snapshot distances can be estimated from complete light-curves, we
sought out SNe~Ia which were observed only one to a few times to
emphasize the unique utility of snapshot distance estimates. An
exception to this was SN 1997I, where we used the two photometry
epochs with the highest signal-to-noise ratio. (If SN 1997I were
observed only twice for this purpose a conscientious observer would
likely have insured such high signal-to-noise ratios with generous
telescope integrations.) Condition 3 is only necessary for the purpose
of measuring the precision of snapshot distance estimates from
deviations from the Hubble law.

Our search was neither exhaustive nor complete.  Our procedure was to
scan the IAU Circulars for SNe~Ia which seemed likely to meet the
three conditions listed above.  After identifying a suitable
candidate, we contacted the observer listed in the Circular to obtain
the data.  In Table~\ref{data_tab} we have listed the data and its
source for a set of seven sparsely observed SNe~Ia.  In all cases the
spectra were calibrated by the listed observer.  All photometry was
calibrated by the authors of this paper and is presented in Appendix
B.

\begin{figure}[p]
\psfig{file=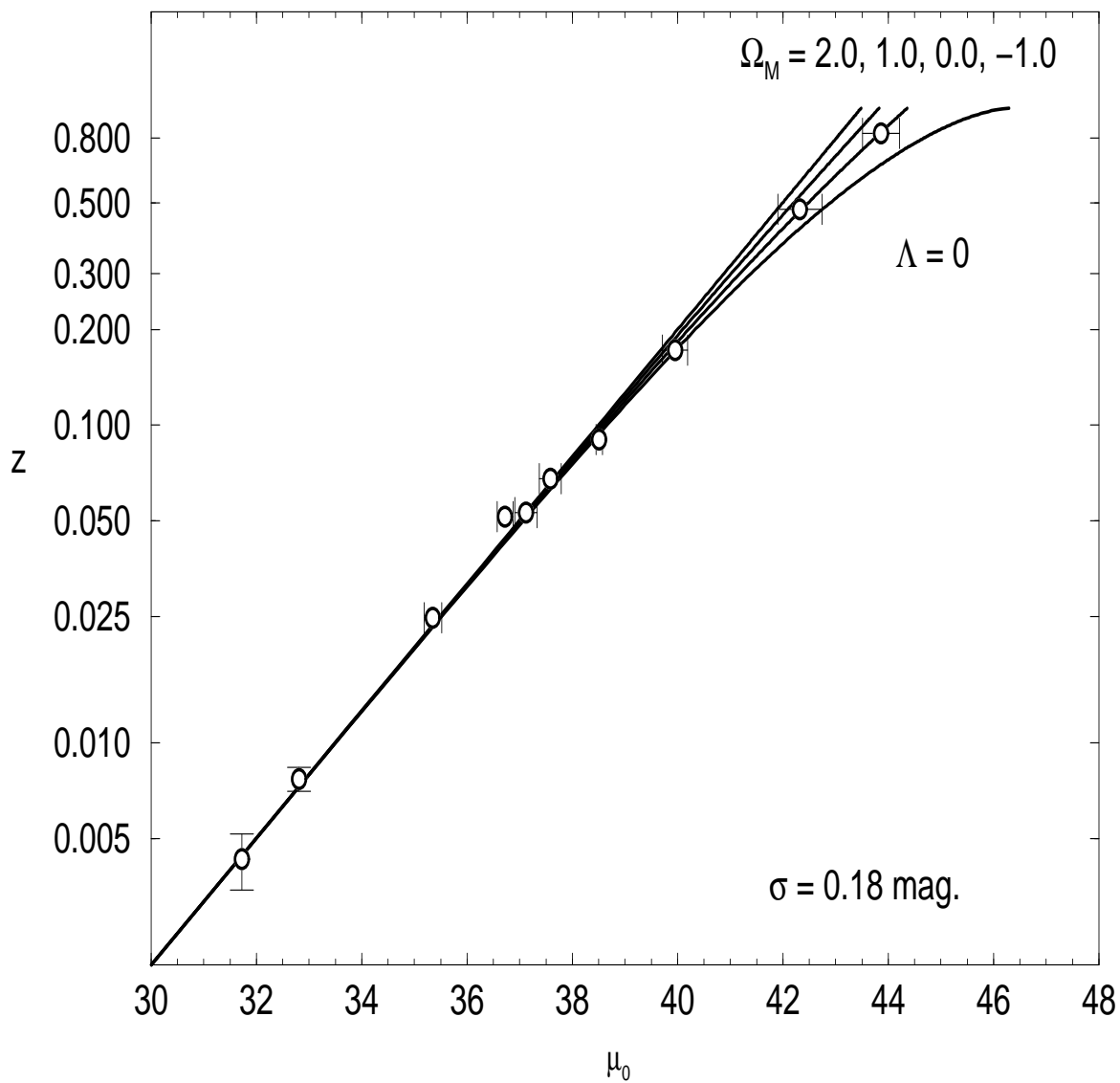,height=7.0in,width=7.in,angle=270}
\caption{Hubble diagram of sparsely observed SNe~Ia with distance
estimates from the snapshot method.  The dispersion is 0.18 mag (for
$z$ $\leq$ 0.2), verifying the estimated and simulated distance
precision of 10\%.  Redshift-magnitude relations for different values
of the density parameter, $\Omega_{\rm M}$, are indicated (assuming no
cosmological constant).\label{hub}}
\end{figure}

We measured the distance to the SNe~Ia listed in Table~\ref{data_tab}
with the method described in \S 2.  Typically, each SN~Ia was observed
one or two times photometrically and spectroscopically.  The redshifts
of the host galaxies and the parameters related to the SNe~Ia distance
estimates are given in Table~\ref{dist_tab}. In all cases we assume a
300 km s$^{-1}$ velocity error to account for velocity flows. The
Hubble diagram constructed from this information is shown in
Figure~\ref{hub}.  Distance errors were calculated by propagating the
effects of perturbing age, luminosity, and photometry measurements by
their uncertainties around their expected values.  The dispersion
around the best fitting Hubble line is 0.18 mag, highly consistent
with the error estimates in \S 3 and the Monte Carlo simulations in \S
3.

\section{Snapshot Distances at High Redshift}

The advantage of the snapshot method over light-curve shape methods is
the dramatic reduction in required observing time for a modest
reduction in distance precision. This advantage can be of great
benefit at redshifts $z \geq 0.3$. In this searchable volume, the
number of SNe~Ia which can be discovered far exceeds the number whose
light-curves can be monitored.

At these redshifts, SNe~Ia are discovered by subtracting past galaxy
images from current ones to detect new objects
\cite[]{perl97,schetal97}.  Candidates are identified as SNe~Ia
through certain characteristics of their spectra \cite[]{bfn93,fil97}.
Continued monitoring of the light-curves is done at great expense of
telescope time and effort, yet such additional supernova data may not
be necessary.

In \S 3 we calculated the uncertainty of a snapshot distance for SNe
Ia within $z \leq 0.1$ to be $\sim$ 10\%.  This result was
independently verified from both Monte Carlo simulations of
well-sampled SNe~Ia and from the Hubble diagram of sparsely sampled
SNe~Ia.  We can repeat this calculation for high-redshift SNe~Ia,
including a larger uncertainty in age estimates and photometry of more
distant supernovae.  \cite{riess_age97} estimate a mean age
uncertainty of 2.0 to 2.5 days from spectra obtained at the Keck-II
telescope of SN 1996bj, a SN~Ia at $z=0.574$.  Images of SN 1996bj
obtained near maximum at the Blanco 4 m telescope at the Cerro Tololo
Inter-American Observatory recorded the brightness of the supernova
with a precision of 5\% \cite[]{schmidt97}. Incorporating these
uncertainties into equation~(\ref{sig2}) results in an expected
snapshot distance uncertainty of 15-20\% or 0.3-0.4 mag (see
Figure~\ref{one}).

As a test of the feasibility of measuring a snapshot distance for SNe
Ia with $z \geq 0.3$ we measured the distance to SN 1995K, a SN~Ia at
$z=0.478$ \cite[]{schetal97} and SN 1997ap at $z=0.83$ \cite[]{97ap}.

An identification spectrum of SN 1995K was obtained at the New
Technology Telescope at ESO by B. Leibundgut and J. Spyromilio on 3
April 1995.  Complete light-curves in $B$ and $V$ were obtained by
\cite{schetal97}.  We have used the identification spectrum and the
two photometry epochs with greatest signal-to-noise ratio (JD 2449807,
2449813) to measure a snapshot distance to SN 1995K.  If SN 1995K had
been sparsely observed for the purpose of measuring its snapshot
distance, the paucity of observations would have merited long
telescope integrations to reach a signal-to-noise ratio at least as
high as that for the observations of JD 2449807 and 244981.  The
spectral feature age of SN 1995K on April 3 was 1$\pm2$ days after
maximum light.  This age is in good agreement with the complete
light-curves for SN 1995K \cite[]{schetal97}. The value of
$\cal{R}$(\ion{Si}{2}) was found to be 0.30, indicating that SN 1995K
was of typical luminosity for most SNe~Ia (see Appendix A). The
measured colors are not redder than those expected for the estimated
age and luminosity, indicating no significant extinction of the
supernova light.  The estimated distance modulus of SN 1995K is
$\mu_0=42.32 \pm 0.41$ mag (see Figure~\ref{hub}).  This snapshot
distance estimate is highly consistent with the more precise MLCS
estimate of $\mu_0=42.40 \pm 0.25$.  Yet, the investment of time to
gather the data for the snapshot estimate was less than 10\% of that
needed for the MLCS estimate.

SN 1997ap was discovered by the Supernova Cosmology Project on 5 March
1997 UT, during a two-night supernova search at the CTIO 4-m
telescope. The supernova light-curve was followed with scheduled
$R$-band and $I$-band imaging at the CTIO, WIYN, ESO 3.6 m, and INT
telescopes, and with spectroscopy at the Keck-II telescope. In
addition, SN 1997ap was followed with scheduled imaging with the {\it
Hubble Space Telescope}. The spectrum and full light-curves can be
seen in \cite{97ap}. Once again, we have used the identification
spectrum and the two photometry epochs with greatest signal-to-noise
ratio (JD 2450547, 2450555) to measure a snapshot distance. The SFA
method \cite[]{riess_age97} was developed to measure the age from
features in the SN~Ia spectrum in the rest wavelength range of 3800 to
6800~\AA.  The substantial redshift of SN 1997ap carries many of these
features out of the optical window, yet brings equally useful features
from the ultraviolet into our window.  To make better use of the
available data, \cite{97ap} have extended the SFA method to use the UV
features at the rest wavelength range of 2750 to 3800~\AA. Spectra
from the International Ultraviolet Explorer \cite[]{iuesne} in
combination with ground-based spectra were used to perform these
calculations.  The March 14 spectrum yields an age of $2 \pm 3$
SN-restframe days ($\sim$4 observer's days) before the supernova's
maximum light in the restframe $B$-band which is consistent with a
supernova age of $1 \pm 2$ SN-restframe days after maximum inferred
for this date from the fit to the light curve.  The value of
$\cal{R}$(\ion{Ca}{2}) was found to be 1.32, which indicates that the
supernova was slightly overluminous ($\Delta=-0.05$ mag). Both the age
and the measurement of the light-curve shape from the spectrum are in
excellent agreement with the results from the full light-curve
analysis as seen in \cite{97ap}. The $U-B$ colors at maximum are
slightly bluer ($E_{U-B}=-0.03 \pm 0.10$ mag) than the expected color
at maximum of $U-B=-0.32$ mag for supernovae with similar light-curve
shapes, indicating that SN 1997ap suffers no extinction [though the
uncertainty on both measurements is large \cite[]{97ap}]. The snapshot
distance modulus of SN 1997ap is $\mu_0=43.86 \pm 0.35$ mag, in exact
agreement with the full light-curve analysis (where the error bars are
only $\pm 0.15$ mag), yet just a small fraction of the available data
were incorporated.

For the snapshot distances in Table 2, we find for a flat universe
($\Omega_{\rm M} + \Omega_{\Lambda} = 1$) that $H_0=63.4 \pm 2$ $\pm
6$ km s$^{-1}$ Mpc$^{-1}$ and $\Omega_{\rm M} = 0.19^{+0.32}_{-0.19}$
with a reduced $\chi^2 = 1.04$. For a $\Omega_{\Lambda} = 0$ universe,
$H_0=62.4 \pm 2$ $\pm 6$ km s$^{-1}$ Mpc$^{-1}$ and $\Omega_{\rm M} =
-0.31^{+0.62}_{-0.36}$ with a reduced $\chi^2 = 1.13$. The former
uncertanties in the Hubble constant reflect the variance in the Hubble
flow; the latter uncertainties result from the uncertainty in the SN
Ia absolute magnitude and the Cepheid distance scale (Feast \& Walker
1987; Kochanek 1997; Madore \& Freedman 1987).  Our results for this
very preliminary study at high redshift are similar to those of
\cite{97ap}, \cite{97hst_hiz} and Riess et al. 1998b which employ full
light-curves; however, we do note a mild disagreement with the earlier
results of \cite{perl97} at the $\sim$ 80\% level. While it is too
early to place strong constraints on the cosmological parameters from
this exploratory study, the results from a small data set of sparsely
observed supernovae are encouraging.

\section{Discussion}
     
In principle, enough information can be garnered from a single
supernova spectrum and photometric epoch to estimate the distance to a
SN~Ia.  In practice, the results of \S 3, 4, and 5 suggest this method
produces distances having a precision of $\sim$ 10\%, with variations
that are a function of the quality of the data and the age of the
supernova.  Depending on the amount of host galaxy contamination, it
may be necessary to obtain spectra and images of the host galaxy after
the supernova has faded.

\begin{figure}[p]
\psfig{file=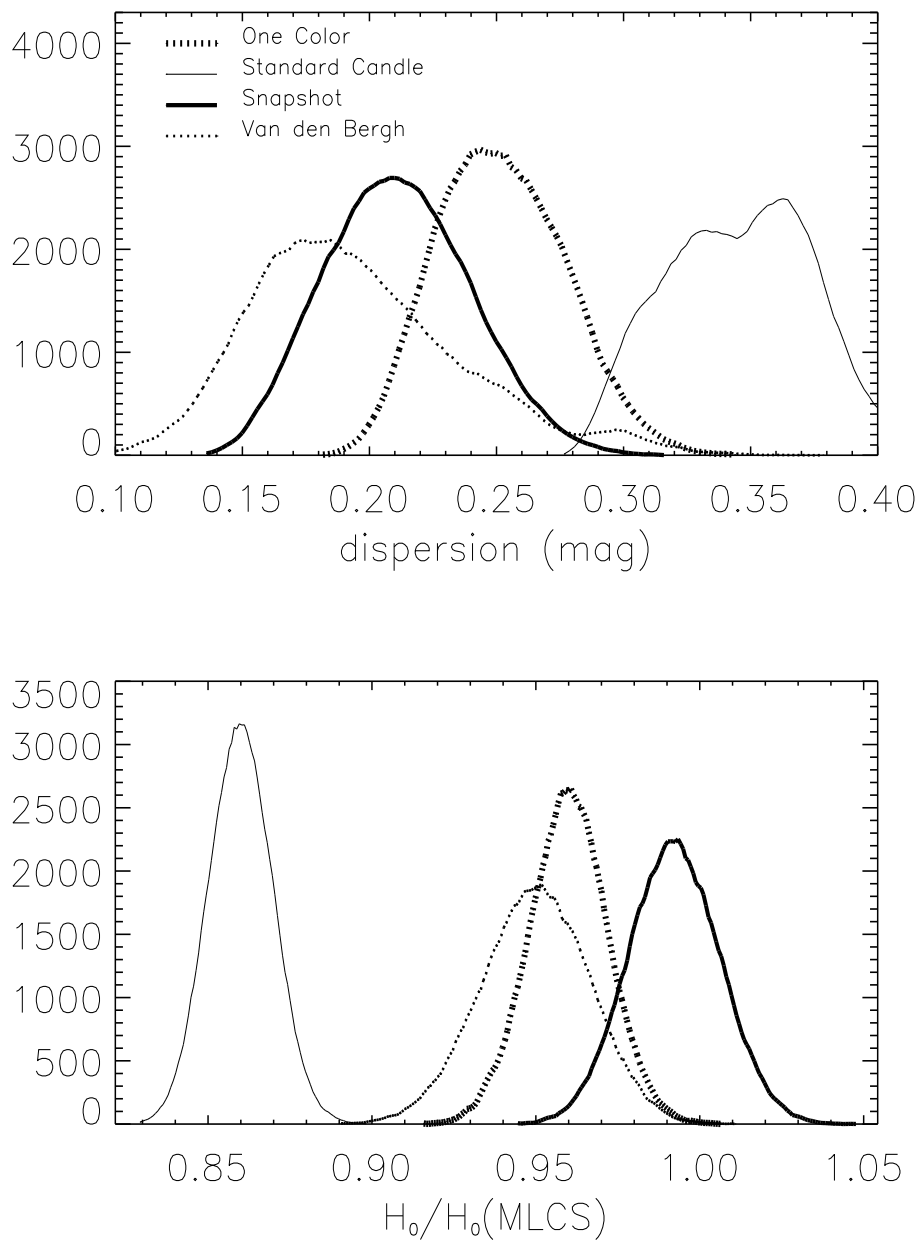,height=7.0in,width=5.5in,angle=0}
\caption{Distributions for SNe~Ia distance methods which employ a
single spectrum and photometric epoch.  A single spectrum and
photometric epoch were randomly selected (not necessarily the same
date) from well-sampled SNe~Ia to measure the distance and the
dispersion from the Hubble law.  Table~\ref{discomp_tab} reviews the
corrections which are included in each distance method.  The snapshot
distance method estimates distances with $\sim$ 10\% precision and no
systematic bias.\label{show}}
\end{figure}

The snapshot distance method employs the same luminosity and
extinction corrections used in the MLCS method of Riess, Press, and
Kirshner 1996a (and more recently updated in Riess et al. 1998b). We
find no significant offset between the distance estimates of the two
methods and only a modest reduction in precision for the snapshot
distance method.  There are three more limited versions of a distance
method using single epoch SN Ia observations which reveal the utility
of luminosity and extinction corrections (see Table~\ref{discomp_tab}
and Figure~\ref{show}).  These variants employ the SFA measurement but
lack the luminosity correction, the extinction correction, or both.

Disregarding individual luminosities and light-curve shapes predicted
by the spectral ratios, we fit homogeneous, fiducial templates to the
photometric epoch.  Further, we discarded our estimate of the
extinction from the color excess.  We measured the resulting
``standard candle'' distances to SNe~Ia using the Monte Carlo
technique described in \S 4.  As seen in Figure~\ref{show}, the
distribution of dispersions has a mean of 0.35 mag, a value consistent
with previous SNe~Ia distance estimates which assume SN Ia light curve
homogeneity and do not correct for extinction
\cite[]{st93,ts95,hametal95,hametal96,rpk95,rpk96,branch_miller,vauetal95}.
These distances are also 15\% {\it greater} in the mean (or smaller in
the implied Hubble constant) than either MLCS or snapshot distances,
consistent with other comparisons of distance estimates which assume
homogeneity instead of heterogeneity of SNe~Ia
\cite[]{ts95,hametal95,hametal96,rpk95,rpk96}.

To simulate the effect of single filter information, we used our
luminosity correction without an extinction correction.  This
procedure results in a mean dispersion of 0.25 mag and distances which
are only 3\% greater than those obtained from MLCS.  This result,
though better than the standard candle method, is still worse than the
complete snapshot procedure.

A final variant is to disregard a luminosity/light-curve shape
correction but maintain an extinction correction from the color
excess.  Such a method using light-curves was proposed by \cite{vdb95}
to account for both intrinsic luminosity differences as well as
absorption by dust.  This method takes advantage of the coincidental
near-agreement between the standard reddening law and the relation
between intrinsic color and luminosity to correct for both extinction
and luminosity differences.  \cite{rpk96} have noted that while both
sources of luminosity variation affect the SN color in the same
direction, the specific ratios of the luminosity difference to color
difference are not precisely the same for extinction and intrinsic SN
Ia variation.  Monte Carlo simulations of this method combined with a
SFA measurement give a mean dispersion of 0.19 mag and distances which
are 5\% greater in the mean than those of the MLCS method.  Despite
the low dispersion of this method, we are suspicious of the distances
it predicts.  The distribution of dispersions obtained from our Monte
Carlo simulation (see Figure~\ref{show} and Table~\ref{discomp_tab})
is more skewed than any other, including an asymmetric ``tail''
encompassing dispersions greater than 0.3 mag.  We believe that for
SNe~Ia with only moderate amounts of extinction or whose luminosities
are similar to those of typical SNe~Ia, this method has merit.  Yet
for very red SNe~Ia, this method can predict distances which are
systematically and considerably in error due to the inability to
distinguish between absorption by dust and intrinsic variation.

The snapshot method predicts distances which agree in the mean with
only a moderate reduction in precision from light-curve shape methods.
Yet because of the greatly diminished expense in data collection, this
method can be more effective for problems which benefit equally well
from a high {\it quantity} of SNe~Ia distances as from the {\it
quality} of those distances.  Two such applications are mapping the
nearby peculiar velocity field and determining the cosmological
parameters which dictate global geometry.

Recent attempts to map the cosmic velocity field with SNe~Ia
\cite[]{riess_beta97} suffer from dilute spatial sampling.  Replete
peculiar velocity maps could reveal the influence of matter
fluctuations and constrain the matter content of the local Universe.
Nearby, many more SNe~Ia are discovered than can be regularly
monitored.  By decreasing the observational requirements of each SN
Ia, it should be possible to increase the sampling of the local
velocity field.  The light-curves employed by \cite{hametal96} and
\cite{rpk96} are typically sampled for 10 to 15 epochs.  The
observational demands increase as the supernova rapidly fades.  With
the telescope time invested in a single set of SN Ia multi-color
light-curves, sufficient data for 15 to 25 SNe~Ia snapshot distances
could be gathered.  Accounting for the inherent distance
uncertainties, telescope time spent collecting snapshot distance data
is 3 to 4 times more efficient than time spent collecting light-curves
for distance estimates.

Efforts to measure the cosmological parameters $\Omega_M$ and
$\Omega_\Lambda$ from distant SNe~Ia could also profit from snapshot
distances.  Systematic searches for SNe~Ia at z $\geq 0.3$ have
yielded a plethora of objects \cite[]{perliau95a,schetal95}.
Combining a new generation of large telescopes with $\sim$ 1 degree
fields of view and multi-fiber spectrometers with the snapshot method
could allow SNe~Ia distances to be gathered in batch at an
unprecedented rate.  At the current rate of discovery, a night spent
searching five 1-degree fields followed by a night collecting spectra
of the candidates with a multi-fiber spectrometer could yield $\sim$
50 SNe~Ia distances \cite[]{schmidt97,rate_96}. Repeating this process
every new moon could yield up to $\sim$ 600 distances a year.  At this
rate of accumulation, it should be possible to convincingly separate
the effects of various sources of energy density on the
redshift-magnitude relation \cite[]{omol_95}.

A more optimal method for measuring SN Ia distances would employ both
the predictive power of SN Ia light and color curve shapes with that
of SN Ia spectra.  Such a method would replace the distinction between
a snapshot distance and a light-curve distance with a distance
estimate which makes the most economical use of all available SN Ia
observations.  Using the tools described in Riess, Press, and Kirshner
1996a, Nugent et al. 1995, Riess et al. 1997, Riess et al. 1998b, and
this paper, such a method appears to be quite feasible.

\bigskip

This work was supported by the NSF through grant AST--9417213 to
A.V.F. and AST-9528899 and AST-9617058 to R.P.K., by the Miller
Institute for Basic Research in Science through a fellowship to
A.G.R., and by the Director, Office of Computational and Technology
Research, Division of Mathematical, Information, and Computational
Sciences of the U.S. DoE under contract number 76SF00098 to P.E.N.
Some of the calculations presented in this paper were performed at the
National Energy Research Supercomputer Center (NERSC), supported by
the U.S. DoE.  We thank Stephan Benetti and George Djorgovski for
allowing us to use their spectra of SNe Ia prior to publication, and
Bruno Leibundgut for suggestions that helped improve this paper.

\appendix

\section{The Correlation of $\Delta$ with $\cal{R}$(\ion{Si}{2}) and
$\cal{R}$(\ion{Ca}{2})}  

Following the prescription by \cite{nugseq95} for measuring
$\cal{R}$(\ion{Si}{2}) and $\cal{R}$(\ion{Ca}{2}) at maximum light, we
present the linear relationships found between these ratios and the
MLCS parameter $\Delta$ (Riess, Press, \& Kirshner 1996a) for a time
sequence of supernovae. The data were binned into three-day intervals
to allow for possible errors in the light-curve estimate of maximum
light ($\sim$ 1.4 days) and to increase the number of data points in
each bin. A complete description of this work is presented in
\cite{phillips97}. In Table~\ref{car_tab} and Table~\ref{sir_tab} we
list the slope ($m$) and intercept ($b$), along with the dispersion
($\sigma_{\Delta}$), for the linear relationship assumed between
[$\cal{R}$(\ion{Si}{2}), $\cal{R}$(\ion{Ca}{2})] and $\Delta$ as
defined by

\bq
\Delta = m*\cal{R} + \rm{b} \pm \sigma_{\Delta}.
\eq

The ratio $\cal{R}$(\ion{Si}{2}) was utilized from 7 days before to 7
days after maximum light and $\cal{R}$(\ion{Ca}{2}) was utilized from
7 days before to 19 days after maximum light. The number of spectra
available earlier than 7 days before maximum light limits our ability
to define the relationships between these ratios and the
luminosity. The technique presented by \cite{nugseq95} for measuring
$\cal{R}$(\ion{Si}{2}) cannot be employed for SNe~Ia older than 7 days
after maximum light because the \ion{Si}{2} trough at 5800~\AA\ loses
its short wavelength boundary.  Likewise the data becomes too sparse
more than 19 days after maximum light, inhibiting our ability to
predict the relationship between $\cal{R}$(\ion{Ca}{2}) and the
luminosity. The scatter in the relationship between
$\cal{R}$(\ion{Si}{2}) and $\Delta$ is $\leq$ 0.25 mag while for
$\cal{R}$(\ion{Ca}{2}) a range of 0.17 $\leq \sigma_{\Delta} \leq$
0.40 mag exists. The \ion{Si}{2} relationship is nearly always an
excellent estimator of intrinsic luminosity while the \ion{Ca}{2}
relationship works best during more limited epochs.  Additional
methods for using the \ion{Ca}{2} feature to measure intrinsic
luminosity are presented in \cite{phillips97}, with an emphasis placed
on how to improve the utility of \ion{Ca}{2}.

\section{Photometry for the $z \leq 0.1$ SNe}

After determining the expected precision of snapshot SNe~Ia distances
to be 10\%, we sought to verify this result on a set of sparsely
observed SNe~Ia.  The selection criteria for such objects are given in
\S 5.  In Table~\ref{data_phot} we list the available photometry. The
photometry was calibrated by the authors of this paper. The methods
used for measuring the brightness of the SN~Ia and its uncertainty are
those of \cite{riessphd}.  Galaxy subtraction was performed on all the
objects followed by point-spread-function (PSF) fitting.  The
exceptions were SN 1997bp and SN1996V, whose brightnesses could be
computed directly from a PSF without significant contamination by
background light.

\clearpage


\begin{deluxetable}{ccccc}
\tablecaption{SN~Ia Observations \label{data_tab}}
\tablehead{
\colhead{} & \multicolumn{2}{c}{Spectra} & \multicolumn{2}{c}{Photometry}\\
\colhead{SN~Ia} & \colhead{Epochs} & \colhead{SN~Ia age} &
\colhead{Epochs} & \colhead{SN~Ia age}
}
\startdata
SN 1994U  & 2 & -7.5$^a$, 8.5$^b$          & 1 &  2.7$^a$          \\
SN 1997bp & 3 & -0.5$^a$, 0.5$^a$, 1.5$^a$ & 1 &  3.5$^a$          \\
SN 1996V  & 1 &  11.7$^a$                  & 1 &  11.7$^a$         \\
SN 1994C  & 2 &  9.5$^a$, 10.5$^c$         & 1 &  14.8$^a$         \\
SN 1995M  & 1 &  9.5$^d$                   & 4 &
                                10.5$^a$, 14.5$^a$, 15.5$^a$, 25.5$^a$\\ 
SN 1995ae & 1 &  8.6$^b$            & 2 &  10.7$^a$, 11.7$^a$      \\
SN 1994B  & 2 & -3.6$^a$, 13.4$^a$  & 2 & -1.3$^a$, -0.3$^a$        \\
SN 1997I$^*$  & 1 & 0.1$^a$             & 2 &  1.0$^a$, 3.0$^a$        \\
SN 1995K$^*$  & 1 & 2.1$^d$             & 2 &  4.2$^d$, 8.2$^d$        \\
SN 1997ap$^*$ & 1 & -2.0$^e$            & 3 & -2.6$^e$, 11.0$^e$, 15.2$^e$\\
\tablecomments{$^{*}$ These supernovae were treated as ``snapshot''
supernovae by picking the two highest signal-to-noise ratio photometric
epochs from their full light-curves.}
\tablerefs{(a) This paper, (b) \cite{bene97}, (c) \cite{george97}, (d)
\cite{schmidt97}, and (e) \cite{97ap}}.
\enddata
\end{deluxetable}

\begin{deluxetable}{ccccccc}
\tablecaption{SN~Ia Parameters\label{dist_tab}}
\tablehead{
\colhead{SN~Ia} & \colhead{log$(cz)$} &
\colhead{JD max} & \colhead{$\Delta$} & \colhead{$A_v$} & 
\colhead{$\mu_0$} & \colhead{$\sigma_{\mu_0}$}\\
\colhead{} & \colhead{(km/s)}  & \colhead{(2400000+)} &
\colhead{(mag)} & \colhead{(mag)} & \colhead{(mag)} & \colhead{(mag)}}
\startdata
SN 1994U  & 3.111 &  9539.0 &  0.03 & 0.70 & 31.72 & 0.10 \\
SN 1997bp & 3.363 & 10546.3 & -0.26 & 0.62 & 32.81 & 0.10 \\
SN 1996V  & 3.870 & 10173.0 &  0.26 & 0.00 & 35.35 & 0.17 \\
SN 1994C  & 4.189 &  9411.0 &  0.81 & 0.00 & 36.72 & 0.15 \\
SN 1995M  & 4.202 &  9822.1 & -0.15 & 0.46 & 37.12 & 0.21 \\
SN 1995ae & 4.308 &  9983.0 &  0.38 & 0.00 & 37.58 & 0.21 \\
SN 1994B  & 4.431 &  9375.1 & -0.02 & 0.38 & 38.51 & 0.10 \\
SN 1997I$^{*}$  & 4.713 & 10461.4 & -0.39 & 0.00 & 39.95 & 0.24 \\
SN 1995K$^{*}$  & 5.157 &  9808.4 &  0.15 & 0.00 & 42.32 & 0.42 \\
SN 1997ap$^{*}$ & 5.396 & 10526.7 &  0.00 & 0.00 & 43.67 & 0.35 \\
\tablecomments{$^{*}$ These supernovae were simulated as ``snapshot''
supernovae by picking the two highest signal-to-noise ratio photometric
epochs from their full light-curves.}
\enddata
\end{deluxetable}

\begin{deluxetable}{cccccc}
\tablecaption{SN~Ia Distance Comparison\label{discomp_tab}}
\tablehead{
\colhead{Correction Method} & \colhead{Luminosity} & \colhead{$A_v$} &
\colhead{$\langle\sigma_{\mu_{0}}\rangle$} & \colhead{Skewness$^{*}$} &
\colhead{H$_0$/H$_{0_{\rm{MLCS}}}$}\\
\colhead{} & \colhead{} & \colhead{(mag)} & \colhead{(mag)} & \colhead{} &
\colhead{}}
\startdata
Standard Candle   & none                                     & none  &
0.35 & 0.22 & 0.86 \\  
One Color         & $\cal{R}($\ion{Si}{2}) and $\cal{R}$(\ion{Ca}{2})& none  &
0.25 & 0.24 & 0.96 \\  
``Van den Bergh'' & none                                     & colors&
0.19 & 0.65 & 0.95 \\
Snapshot          & $\cal{R}$(\ion{Si}{2}) and $\cal{R}$(\ion{Ca}{2})& colors&
0.21 & 0.20 & 0.99 \\ 
\enddata
\tablecomments{$^{*}$ Skewness is defined to be$ \frac{\langle (\mu_0 -
\bar{\mu}_0)^{3} \rangle}{\sigma_{\mu_{0}}^{3}}$}  
\end{deluxetable}

\begin{deluxetable}{cccccccc}
\tablecaption{\label{sir_tab} The relationship between $\Delta$ and
$\cal{R}$(\ion{Si}{2}) $^a$}.   
\tablehead{
\colhead{Phase} & \colhead{$m$} & \colhead{$b$} &
\colhead{$\sigma_{\Delta}$} \\
\colhead{(days)} & \colhead{(mag)} & \colhead{(mag)} & \colhead{(mag)}}
\small
\startdata
-7 & 1.73 & -0.32 & 0.24 \\
-6 & 1.58 & -0.34 & 0.18 \\
-5 & 1.54 & -0.33 & 0.17 \\
-4 & 1.85 & -0.40 & 0.20 \\
-3 & 1.89 & -0.42 & 0.19 \\
-2 & 2.30 & -0.58 & 0.22 \\
-1 & 2.34 & -0.58 & 0.21 \\
 0 & 2.33 & -0.55 & 0.22 \\ 
 1 & 2.20 & -0.51 & 0.21 \\
 2 & 2.34 & -0.54 & 0.21 \\
 3 & 2.07 & -0.45 & 0.19 \\
 4 & 2.06 & -0.46 & 0.20 \\
 5 & 1.90 & -0.47 & 0.22 \\
 6 & 1.13 & -0.28 & 0.23 \\
 7 & 0.67 & -0.15 & 0.25 \\
\enddata
\tablecomments{$^{a}$ The equation $\Delta=m*\cal{R}$(\ion{Si}{2})$ + b
\pm \sigma_{\Delta}$ was used to derive the relationship between
$\Delta$ and $\cal{R}$(\ion{Si}{2}).}  
\end{deluxetable}

\begin{deluxetable}{cccc}
\tablecaption{\label{car_tab} The relationship between $\Delta$ and
$\cal{R}$(\ion{Ca}{2})$^a$}.  
\tablehead{
\colhead{Phase} & \colhead{$m$} & \colhead{$b$} &
\colhead{$\sigma_{\Delta}$}\\
\colhead{(days)} & \colhead{(mag)} & \colhead{(mag)} & \colhead{(mag)} 
}
\startdata
-7 & 0.31 & -0.37 & 0.36  \\
-6 & 0.33 & -0.39 & 0.28  \\
-5 & 0.33 & -0.43 & 0.29  \\
-4 & 0.21 & -0.23 & 0.27  \\
-3 & 0.20 & -0.21 & 0.23  \\
-2 & 0.82 & -1.14 & 0.30  \\
-1 & 0.87 & -1.23 & 0.32  \\
 0 & 0.65 & -0.79 & 0.37  \\
 1 & 0.69 & -0.79 & 0.37  \\
 2 & 0.72 & -0.84 & 0.36  \\
 3 & 0.44 & -0.45 & 0.30  \\
 4 & 0.46 & -0.47 & 0.30  \\
 5 & 0.54 & -0.66 & 0.28  \\
 6 & 0.35 & -0.47 & 0.19  \\
 7 & 0.40 & -0.55 & 0.21  \\
 8 & 0.48 & -0.71 & 0.23  \\
 9 & 0.44 & -0.70 & 0.17  \\ \tablebreak
10 & 1.08 & -1.52 & 0.32  \\
11 & 1.42 & -1.98 & 0.26  \\
12 & 1.23 & -1.74 & 0.35  \\
13 & 1.34 & -1.93 & 0.39  \\
14 & 1.12 & -1.62 & 0.40  \\
15 & 0.94 & -1.39 & 0.31  \\
16 & 0.92 & -1.35 & 0.31  \\
17 & 1.03 & -1.53 & 0.25  \\
18 & 0.96 & -1.37 & 0.22  \\
19 & 0.59 & -0.81 & 0.24  \\
\enddata
\tablecomments{$^{a}$ The equation $\Delta=m*\cal{R}$(\ion{Ca}{2})$ + b
\pm \sigma_{\Delta}$ was used to derive the relationship between
$\Delta$ and $\cal{R}$(\ion{Ca}{2}).}
\end{deluxetable}

\begin{deluxetable}{cccccc}
\tablecaption{SN~Ia Photometry \label{data_phot}}
\tablehead{
\colhead{SN~Ia} & \colhead{JD} & \colhead{$B$} &
\colhead{$V$} & \colhead{$R$} & \colhead{$I$}\\
\colhead{} & \colhead{(2400000+)} & \colhead{(mag)} & \colhead{(mag)}
& \colhead{(mag)} & \colhead{(mag)}  
}
\startdata
SN 1994U & 9541.7 & 13.46(0.03) & 13.07(0.02) & 12.86(0.02) & 13.22(0.04) \\
SN 1997bp & 10549.8 & 14.03(0.03) & 13.79(0.03) & 13.80(0.03) & 14.08(0.04) \\
SN 1996V & 10184.7 & 16.97(0.03) & 16.62(0.03) & 16.69(0.03) & 17.08(0.04) \\
SN 1994C & 9425.8 & 19.95(0.06) & 18.94(0.05) & 18.51(0.02) & 18.15(0.08) \\
SN 1995M & 9832.6 &     18.68(0.04) & 18.28(0.02) & --- & --- \\
     &    9833.6 & --- & 18.33(0.02) & --- & --- \\
     &     9836.7 &     19.15(0.04) & 18.50 (0.02) & --- & --- \\
     &     9837.6 &     19.20(0.04) & 18.54 (0.02) & --- & --- \\
     &     9843.7 & --- & 18.90(0.02) & --- & --- \\
     &     9847.6 &     20.28(0.04) & --- & --- & --- \\
SN 1995ae & 9993.7 & 19.28(0.05) & 18.92(0.03) & 18.78(0.04) &  19.11(0.06) \\
          & 9994.7 & 19.34(0.07) & 18.95(0.03) & 18.74(0.04) & 19.48(0.15) \\
SN 1994B & 9373.8 & 19.62(0.05) & 19.43(0.03) & 19.08(0.04) & 19.68(0.06) \\
         & 9374.8 & 19.67(0.04) & 19.50(0.04) & 19.01(0.04) & 19.57(0.07) \\
SN 1997I & 10462.6 & --- & 20.55(0.03) & --- & --- \\
         & 10464.8 & --- & 20.53(0.03) & 20.70(0.03) & --- \\
         & 10480.7 & --- & --- & 21.31(0.03) & --- \\
\enddata
\end{deluxetable}

\end{document}